\documentclass[12pt]{iopart}

\usepackage{iopams}  
\usepackage{color,graphicx}
\usepackage{epstopdf}

\def \Cs {$^{133}$Cs }
\def  \Hs {$H_{\rm sat}$}
\def  \To {$T_1^{-1}$}
\def  \Tt {$T_2^{-1}$}
\def\CsCl {Cs$_2$CuCl$_4$ } 
\def\CsE {Cs$_2$CuCl$_4$}

\begin{document}

\setcounter{figure}{0}
   
\title[NMR study of Cs$_2$CuCl$_4$]{$^{133}$Cs NMR investigation
  of 2D frustrated Heisenberg antiferromagnet, Cs$_2$CuCl$_4$}

\author{M. -A. Vachon, W. Kundhikanjana, A. Straub, and V. F. Mitrovi{\'c}}
\address{Department of Physics, Brown University, Providence, RI 02912, U.S.A. }

 \author{A. P. Reyes and P. Kuhns}
  \address{National High Magnetic Field Laboratory, Tallahassee, FL 32310, U.S.A.}
  \author{R. Coldea}
\address{Department of Physics, University of Bristol, Bristol BS8 1TL, UK}
  \author{Z. Tylczynski}
 \address{Institute of Physics, Adam Mickiewicz University,  Umultowska 85, 61-614 Poznan, Poland}

\date{\today}


\begin{abstract}
We report $^{133}$Cs nuclear magnetic resonance (NMR) measurements on
the 2D frustrated Heisenberg antiferromagnet Cs$_2$CuCl$_4$ down to
$2$\,K and up to $15$\,T.  We    show that $^{133}$Cs NMR is a good probe
of the magnetic degrees of freedom in this material. 
Cu spin degrees of freedom are sensed through a strong anisotropic 
hyperfine coupling.   
The spin excitation gap opens above the critical saturation field.  The gap value was  determined 
from the activation energy  of the nuclear spin-lattice relaxation rate 
in a magnetic field applied parallel to the Cu chains ($\hat{b}$ axis). 
The values of the $g$-factor and the saturation field are consistent with the neutron-scattering 
and magnetization results. 
The measurements of the spin-spin relaxation time  are exploited to show that no structural changes 
occur down to the lowest temperatures investigated. 
\end{abstract}

\maketitle


\section{Introduction}

Two dimensional quantum antiferromagnets on geometrically-frustrated lattices have been theoretically proposed to show strongly correlated physics and possibly   spin liquid states with no long-range magnetic order and fractional spin excitations (for a review see for example Ref. \cite{Misguich:2002}).  In this respect 
the quasi-two-dimensional insulating 
spin-1/2 antiferromagnet \CsCl with spins 
on an anisotropic triangular lattice has 
recently attracted much experimental \cite{Coldea:2001,Coldea:2003}
and theoretical interest
\cite{Chung:2003,Chung:2001,Zhou:2002, Isakov:2005, Alicea:2005, Yunoki:2006, Veillette:2005, Dalidovich:2006}.  Neutron scattering measurements have revealed several 
unusual features in the spin dynamics that have been interpreted in terms of proximity 
to a spin liquid state \cite{Chung:2003,Chung:2001, 
Zhou:2002, Isakov:2005, Alicea:2005, Yunoki:2006}. 
The question of the existence of a gap is  central to the issue of the spin-liquid 
classification \cite{Misguich:2002}. 
 NMR as a low-energy probe of the spin degrees of freedom  could potentially provide  direct information about the low-energy spin dynamics and the presence of a small spin gap in the field induced phases.   
  
 Here we report on extensive \Cs NMR measurements in single crystals of \CsCl in   fields up to 
 \mbox{15 T},  applied parallel and perpendicular to the magnetic planes, and at 
 temperatures down to 2 K. 
 Both static and dynamic  measurements reveal that  \Cs NMR is   a sensitive probe of the magnetism of Cu ions. 
 The strength of the Cs hyperfine coupling  is deduced. 
 From the activation energy of the nuclear spin-lattice 
relaxation rate we confirm that a Zeeman gap opens at high fields above the ferromagnetic saturation field, \Hs. The gap value  is    consistent with the gap determined by neutron scattering measurements. 
 Exploiting the spin-spin relaxation time  measurements, we   show that no structural changes 
occur  down to the lowest temperatures investigated. 

The rest of the paper 
is organized as follows. The crystal structure and magnetic properties of Cs$_2$CuCl$_4$ are described in Sec. 2. In Sec. 3, we present  measurements of the spin shift and use these to extract hyperfine coupling constants. In Sec. 4,    measurements of the nuclear spin-lattice relaxation rate  \To  are used to determine spin gap values. 
In addition, we discuss a special technique that was employed to measure
the rate. 
Finally, the spin-spin relaxation rates are discussed in  Sec. 5. These measurements are used to deduce 
quadrupole coupling
  parameters at low temperatures.

 
\section{Crystal Structure, Magnetism, and  Experimental Details}
\label{Crystal}

\begin{figure}[t]
\begin{center}
\centerline{\includegraphics[scale=1.23]{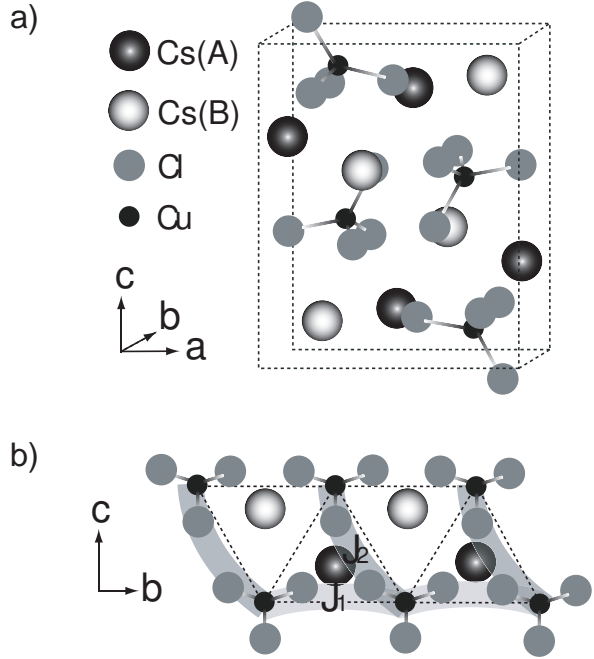}}
\caption{\label{unitcell} {\bf (a)} Unit cell of Cs$_2$CuCl$_4$ with the
  Cu$^{2+}$ ions displayed as small black  spheres in the center of the tetrahedra formed by    the Cl$^{-}$ ions.  The chains are
  along the $\hat{b}$ axis and the planes are in the ($\hat{b}\hat{c}$) 
  plane. {\bf (b)} Triangular magnetic lattice with exchange couplings
  of $J_1=0.374$\,meV and $J_2=0.128$\,meV.}   
\end{center}
\end{figure}

The crystal structure of \CsCl is orthorhombic with space group $Pnma$ and lattice
parameters \mbox{$a=9.65$\,\AA}, \mbox{$b=7.48$\,\AA}, and \mbox{$c=12.35$\,\AA} at
\mbox{$T=0.3$ K} \cite{Bailleul:1991}. 
As displayed in \mbox{Fig. \ref{unitcell}(a)}, each unit cell 
consists of four CuCl$^{2-}_4$ tetrahedra and eight Cs atoms.
The tetrahedra form a linear chain in the $\hat{b}$ direction with two Cl
between each Cu. The chains are stacked together along the $\hat{c}$ direction,
displaced by $b/2$ with respect to each other.   The 
planes are stacked together along the $\hat{a}$ direction.  
In this structure there are two 
inequivalent cesium nuclear sites, labeled as Cs(A) and Cs(B). 
 The Cs(A) atoms,  surrounded by eleven Cl at a mean distance of
$3.79$\,\AA\,   are located in the center of the Cu triangles
close to the plane.
The Cs(B) atoms, surrounded by nine Cl at a mean distance of $3.59$\,\AA,  are situated between the planes and closer to a side of the Cu triangles \cite{Mcginnety:1972}.

%
%
The material is an insulator with each magnetic Cu$^{2+}$ ion carrying 
spin $S=1/2$. The superexchange routes are mediated by 2 nonmagnetic Cl$^-$ ions
and form a frustrated triangular lattice in the $(\hat{b}\hat{c})$
plane, see \mbox{Fig. \ref{unitcell}(b)}. The anisotropic coupling exchange constants are
\mbox{$J_1=0.374$\,meV} along the chains ($\hat{b}$ axis) and
\mbox{$J_2=0.128$\,meV} between the chains ($\hat{c}$ axis)\,\cite{Coldea:1996a}. 
There is   a small interplane coupling \mbox{$J_3=0.017$\,meV} along the
$\hat{a}$ direction that leads to long range order  below \mbox{$T_N=0.62$\,K} in 
zero applied field \cite{Coldea:2001}. The order is in the form of a spiral because of the frustrated in-plane couplings \cite{Coldea:2001}. 
Finally,    the  strength of the Dzyaloshinskii-Moriya term
between the chains is found to be  \mbox{$D_a=0.020(2)$\,meV} 
\cite{Coldea:2002}. Because of
the frustrated spin geometry on a triangular lattice,  a  very rich low temperature phase diagram is 
observed \cite{Tokiwa:2005, VachonUP}. 

We have used a solution-growth  single crystal of Cs$_2$CuCl$_4$ (cut to \mbox{$3.2$ x $6$ x $1.9$\,mm)}. 
The measurements were done at the National High Magnetic Field Laboratory (NHMFL) in
Tallahasee, Florida,  using a high homogeneity 15 T sweepable NMR magnet. 
The temperature control was provided by 
$^4$He variable temperature  insert.
The NMR data were recorded using a 
state-of-the-art homemade NMR spectrometer. 
NMR spectra were obtained, at each value of the applied field,  from the sum of spin-echo Fourier transforms recorded at each 
50 (or 100) KHz intervals. The shift was obtained from the first spectral moment   using a 
gyromagnetic ratio  of the bare \Cs nucleus,   \mbox{5.5844 MHz/T}.

 
\section{\Cs Shift Measurements}
\label{shift}

\begin{figure}[t]
\begin{center}
\includegraphics[width=8cm]{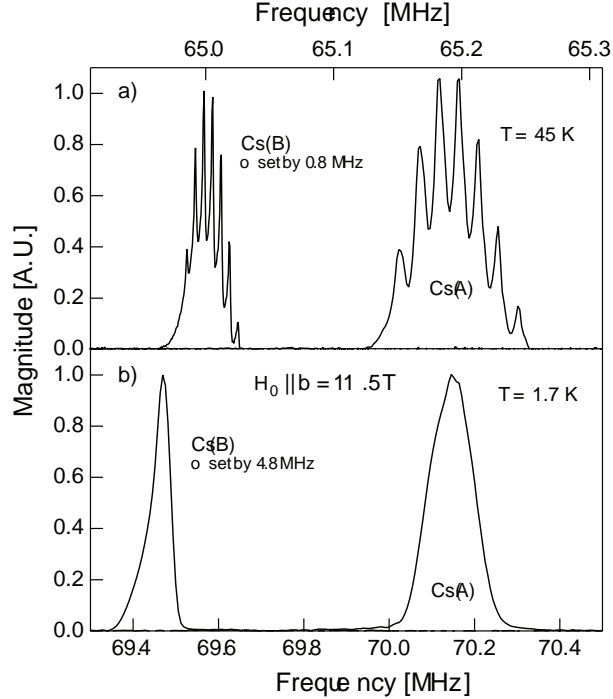}
\caption{\label{spectrum} $^{133}$Cs NMR resonance frequencies at
  $T=45$\,K {\bf (a)} and $T=1.7$\,K {\bf (b)} at $H_0\,||\,b=11.5$\,T. The quadrupole
  splitting   dissapears at low temperatures due to magnetic broadening of
  spectra.} 
\end{center}
\end{figure}

The \Cs nuclear spin is $I=7/2$ and since two Cs sites are both in noncubic environments \cite{Hartman:1968, Klaassen:1973,Lim:2004} quadrupolar effects are relevant. 
Therefore, seven distinct NMR satellite lines are observed for each Cs site, as shown in 
  \mbox{Fig. \ref{spectrum}(a)}. 
To the  first order, the  NMR spectrum  consist of the following
frequencies \cite{Abragam:1961}
\begin{eqnarray}
\label{eq1}
  \displaystyle 
\omega_{\rm NMR}  &&= \, \gamma (1 + K)H_0 \, + 
   \nonumber \\
  \displaystyle   &
&\, \,\omega_Q( m -1/2)(3
\cos^2 \theta -1 + \eta \sin^2 \theta \cos 2 \phi),
\end{eqnarray}
where $\gamma$ is the nuclear gyromagnetic ratio of   $^{133}$Cs, and  $K$ is
the net shift along the applied field direction and is  unique for each site.
The second term accounts for the quadrupole interaction, i.e., seven peaks displayed in \mbox{Fig. \ref{spectrum}(a)}, for each 
$m \leftrightarrow  m \pm 1$ transition. Here, $\nu_Q = 2 \pi \omega_Q =
3e^2qQ/(h2I(I-1))$,    $Q$ is
the nuclear quadrupole moment,    $\eta$  the asymmetry parameter, and $\theta$ and $\phi$ 
are the spherical angles between the applied field and the principal axis of the electric field gradient (EFG). 
The EFG tensors are nearly symmetric for both sites, but their principal axes do not coincide. 
The quadrupole frequencies were deduced from the   quadrupole splitting and 
are consistent with previously reported high temperature values   \cite{Hartman:1968, Klaassen:1973,Lim:2004}. 
We find that: 

\begin{center}
\begin{tabular}{cc}
$\left(\omega_Q \right)_a^{\rm CsA} = 27(2)$\, kHz,  &  $\left(\omega_Q \right)_a^{\rm CsB} = 8.5(5)$\, kHz, \\
\\
$\left(\omega_Q \right)_b^{\rm CsA} = 16(1)$\, kHz,  &  $\left(\omega_Q \right)_b^{\rm CsB} = 6.6(2)$\, kHz. \\
\end{tabular}
\end{center}
In our samples, these values are temperature and  field-independent. 
Nonetheless, below $T \sim 20$ K, quadrupolar satellites cannot be discerned due to the magnetic line broadening (see \mbox{Fig. \ref{spectrum}(b)}). Therefore, at  
 low temperatures  the quadrupole frequencies were deduced from 
the beats   in the spin-spin relaxation time profile (see \mbox{Sec. \ref{T2}}).

The net anisotropic shift consists of several contributions given by

\begin{equation}
\label{Kcw}
K_{\alpha}(T) = K^{\rm orbit}_{\alpha} + K^{\rm quad}_{\alpha} + K^{\rm hf}_{\alpha}(T),
\end{equation}
where $\alpha$ refers to the direction of the applied field
with respect to crystalline axis. $K^{\rm orbit}_{\alpha}$ is the orbital shift and is
temperature-independent and $K^{\rm quad}_{\alpha}$ is the contribution from
the quadrupole interaction.  $K^{\rm hf}_{\alpha}(T)$ is
the hyperfine shift and  is given by: 

\begin{equation}\label{Hhf}
 K^{\rm hf}_{\alpha}(T)= \mu_B \sum_i g_{i} A_{\alpha \,i} \langle S_{i}
 \rangle / H_0,
\end{equation}
where $\mu_B$ is the Bohr magneton, $i=\{a,b,c \}$,  \mbox{$g_{i}$}  the anisotropic g-factor, $A_{\alpha\, i}$ the anisotropic hyperfine tensor, and $\langle
S_{i} \rangle$ is the local spin density of Cu spin in the $i$ direction.

\begin{figure}[t]
\centerline{\includegraphics[scale=0.5]{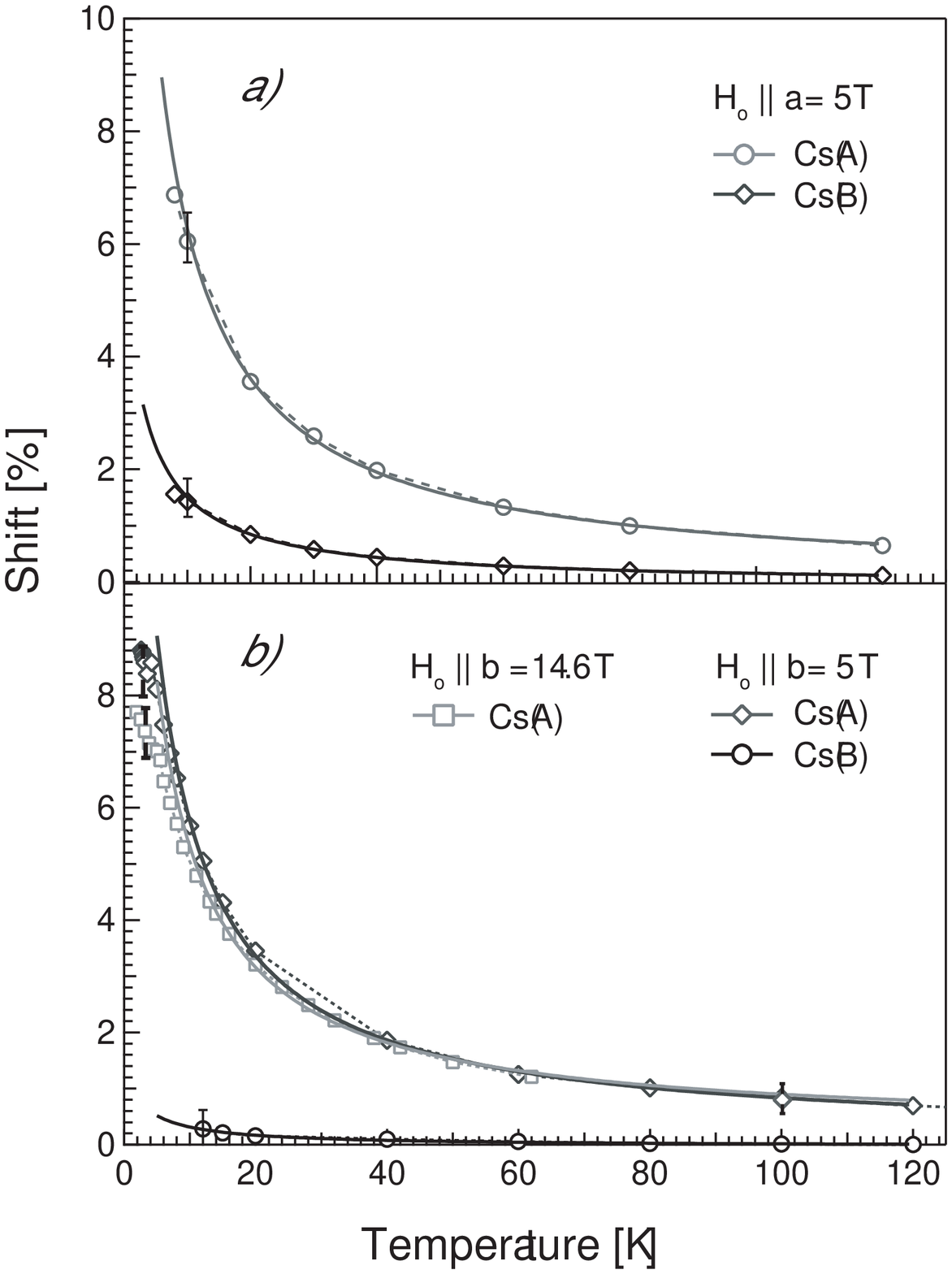}}
\caption{\label{KS} Net shift as a function of temperature at 
$H_0= 5$\,T  field applied parallel to  $\hat{a}$ axis, {\bf (a)}, and parallel to  $\hat{b}$ axis, {\bf (b)}. 
The dashed lines are guide to the eye. 
Typical error bars are shown. 
The solid lines are fit to \mbox{Eq. (\ref{FitCW})}. 
The net shift is nearly isotropic for Cs(A), but highly anisotropic for Cs(B), 
its   value for  $H_0 || \hat{a}$ being more than twice the value for $H_0 || \hat{b}$.}   
\end{figure}

The magnitude of each contribution to the
shift is  determined. The temperature dependence of  the net shift of Cs(A) and Cs(B) sites
   at $H_0=5$\,T   is shown in  \mbox{Fig. \ref{KS}}. 
Furthermore,  the net shift of Cs(A) site at 
$H_0\,||\, \hat b = 14.6$\,T  is displayed. It is evident that within the error bars 
  the shift  is    field independent.
Although  all the curves display a similar temperature dependence, we note a net difference 
between the shift of Cs(A) and that of Cs(B). 
At low temperatures, this difference depends strongly on the direction of the field. 
For a field applied along the crystalline $\hat b$  axis, 
$K_b^{Cs(A)}$ equals approximately  $8 K_b^{Cs(B)}$, whereas $K_a^{Cs(A)} \sim 4 K_a^{Cs(B)}$
 for field applied along the $\hat{a}$ axis.
This difference cannot be accounted for by a dipole-dipole
interaction since the position of Cs(A) and Cs(B) are approximately  the same
with respect to the Cu atom. Indeed, calculations of the dipolar fields confirm this supposition. 
Therefore, the difference is caused, most likely, by a strong anisotropic transferred  
hyperfine interaction with the Cu spins via the Cl atoms. 
 
The temperature dependence of the shift can be expressed in the following form, 
\begin{equation}
\label{FitCW}
K(T) = K^{orbit} + \frac{C}{T  - \theta_{cw}},
\end{equation}
where $K^{orbit}$ is the temperature independent orbital shift, $C$ a constant,  
and $\theta_{cw}$ is the Curie-Weiss temperature.
High-temperature series expansion  \cite{Zheng:2005} 
   implies that $\theta_{cw}= -(J_2 + J_1/2)$, 
which  equals   $3.63$\,K for Cs$_2$CuCl$_4$.  This value of $\theta_{cw}$ provides a good fit for the temperature dependence of  the data for \mbox{$T \gtrsim 8$ K} as illustrated in \mbox{Fig. \ref{KS}}.
If we allow $\theta_{cw}$ to be a free fitting parameter, we obtain  small positive values 
for the Curie-Weiss temperature for shift data for both Cs sites and all values of the applied magnetic field. The result is  in agreement with the predictions of  high-temperature series expansion
 calculation. 
This demonstrates  that the fluctuations are antiferromagnetic-like for temperatures 
higher than  \mbox{$\sim  8$\,K} at all values of the applied field.
Deviations from the Curie-Weiss temperature dependence  are observed below $T \sim 8$\,K. 
This is indeed expected as a precursor  of   a short range order phase that 
develops in the vicinity of  $T \sim 2.5$\,K for all direction of the applied field \cite{Tokiwa:2005}. 

\begin{figure}[t]
\centerline{\includegraphics[scale=0.5]{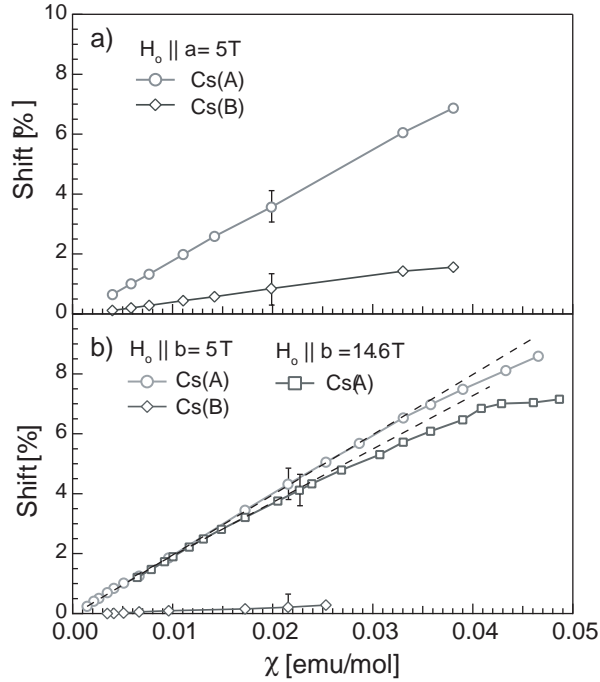}}

\caption{\label{CJplot} $^{133}$Cs Clogston-Jaccarino plots at $H_0=5$\,T.
{\bf (a)} For field applied  along $\hat{a}$. {\bf (b)} For field 
  applied  along $\hat{b}$. Black dashed line are linear fit to the data.} 
\end{figure}

In   \mbox{Fig. \ref{CJplot}} the Clogston-Jaccarino plots \cite{Clogston:1964}, used to 
determine hyperfine coupling constants and orbital shifts,  are shown.
More precisely, the slope of the graph is related to the strength of the hyperfine coupling, while the zero intercept gives the orbital shift. 
Bulk susceptibility data can be fit to    the anisotropic Curie-Weiss
law \mbox{$\chi = C/(T-\theta)$ with $C = N_Ag^2 \mu_B^2S(S+1)/3k_B$} 
 giving  \mbox{$\theta= 4.0 \pm 0.2 $ K},   for \mbox{$T > 20 $ K}, and  g-factors of 
$g_a=2.27$,  $g_b=2.11$, $g_c=2.36$
 along  the $\hat{a}$, $\hat{b}$ and $\hat{c}$ axes, respectively    \cite{Tokiwa:2005}.
Linear behavior, of the form
$K_{\alpha}= K_{\alpha}^{ \rm orb} + m_{\alpha}\chi$, is found for $T>8$\,K. 
Assuming that $K_{\alpha}^{\rm hf}$ equals to  $g_{\alpha}  A_{\alpha \, \alpha} \chi$, we infer
$A_{ \alpha \,\alpha} = m_{\alpha} /g_{\alpha}$ in the units of \mbox{[T/$\mu_B$]}: 
\begin{center}
\begin{tabular}{cc}
$A_a^{\rm Cs(A)} = 0.83(2)$\,T/$\mu_B$,  &  $A_b^{\rm Cs(A)}=0.96(2)$\,T/$\mu_B$, \\
 \\
$A_a^{\rm Cs(B)} = 0.20(2)$\,T/$\mu_B$,  &  $A_b^{\rm Cs(B)}=0.06(1)$\,T/$\mu_B$.
\end{tabular}
\end{center}
The results are obtained at $H_0=5$\,T. 
Moreover, the measurements at \mbox{$H_0=5$\,T} give \mbox{$K^{orbit}_{\alpha}= -0.03(5) \%$}, but
measurements at higher field \mbox{($H_0>12$\,T)} give  
  \mbox{$K^{orbit}_{\alpha}= 0.3(1)\%$}.  In either case,     the orbital shift is negligible compared to the dominant magnetic shift at low temperatures.

From these results, we note that the hyperfine coupling is 
stronger at  Cs(A) than at Cs(B). 
What is more, a more pronounced anisotropy is seen at  Cs(B): the hyperfine coupling is 
more than three time stronger for a field applied along the $\hat{a}$ direction than 
 when  $H_0$ is parallel to $\hat{b}$.
The anisotropy  for Cs(A) site is  weak.  

The values of the hyperfine coupling constants are in agreement  with the internal field measured at full polarization.
Since at low temperature all the spins are polarized along the applied field for 
\mbox{$H_0 >H_{\rm sat}$}, we can assume
 that \mbox{$H^{\rm{int}}_{\alpha} = \mu_{\alpha}  A_{\alpha,\alpha}$}. By taking the magnetic moment at full polarization to be \mbox{$\mu=(1.104\,\mu_B$,  $1.04\,\mu_B$, $1.17\,\mu_B)$} \cite{Coldea:1996a, Tokiwa:2005}, and the $H^{\rm{int}}$ values as described in \cite{NotPolField} and \mbox{Sec. \ref{T1}}, we obtain:
\begin{center}
\begin{tabular}{cc}
$A_a^{\rm Cs(A)} = 1.02$\,T/$\mu_B$,  & $A_a^{\rm Cs(B)} = 0.25$\,T/$\mu_B$,\\
 $A_b^{\rm Cs(A)}=1.11$\,T/$\mu_B$,  &  $A_b^{\rm Cs(B)}=0.09$\,T/$\mu_B$,\\
 $A_c^{\rm Cs(A)}=1.24$\,T/$\mu_B$,  &  $A_c^{\rm Cs(B)}=0.074$\,T/$\mu_B$.
\end{tabular}
\end{center}
These values are somewhat higher than the values found using the  shift measurements, but the ratios  
are the same.
We emphasize  that these values remain, to a good approximation,
independent of the applied field and temperature.

The small  quadrupole interaction  of  the order of tenths of kHz leads  to a 
negligible quadrupolar shift  of the order of 
\mbox{$\sim {\omega_Q^2/\omega_0}$}, i.e. 0.05 \% of  the magnetic hyperfine shift. Moreover, since the orbital shifts are small, 
the major contribution to the shift comes from the hyperfine
interaction between the Cs and  Cu spins. Therefore, NMR on $^{133}$Cs
is a good probe of magnetic degrees of freedom, and the local spin
density is accessible via the measurement of the hyperfine fields, see \mbox{Eq. (\ref{Hhf})}.

We were  unable  to detect the Cu NMR signal due to fast electronic  spin fluctuations
 that  lead to very fast nuclear spin relaxation  times of the order of \mbox{$\hbar/J \sim 10^{-13}$ s}.
However, 
 the chlorine signal   was detected at low temperatures at  applied fields stronger than $12$\,T. %
  It is evident that the  Cl nuclear spins, situated along the exchange path, 
 sense strong electronic spin fluctuations. This   leads to an extremely  
    fast spin-spin relaxation rate that inhibits  detection of the signal.
  As the field increases, the fluctuations are
damped due to the opening of the spin gap. Consequently, the relaxation rate is
slowed down and the signal becomes detectable. An average internal hyperfine field 
 sensed by the Cl  is $H_{hf}\sim 4.6$\,T. This is large  compared to 
 internal field ($H_{hf}\sim 1$\,T) sensed by  the Cs(A).


\section{ Spin-Lattice Relaxation Time and  the Spin Gap}
\label{T1}
At low temperature ($T=0.2$\,K),  a fully spin-polarized state is achieved when the applied field exceeds
the saturation field, $H_{sat} \simeq 8.44$\,T along $\hat{a}$  and $H_{sat}  \simeq 8.5$\,T along $\hat{b}$ 
\cite{Coldea:2002, Tokiwa:2005}. In this regime, the ground
state is ferromagnetic    and excitations are gapped 
magnons with a gap value that increases with the increasing field. 
The field dependence of the gap is given by
 \begin{equation}
 \label{DelH}
  \Delta_{\rm H} = g \mu_B (H_0 - H_{\rm sat}), \; \; \; \; {\rm for} \; H_0 > H_{\rm sat},
\end{equation}
and $E_{\rm sat}\equiv g \mu_B H_{\rm sat}$.
This state has been extensively studied by neutron scattering 
\cite{Coldea:2002}. The spin-lattice relaxation rate ($T_1^{-1}$) measurements
provide an alternative method to probe this ferromagnetic region, 
 potentially   confirm the results obtained by neutron 
experiments, and thus demonstrate that \Cs NMR is a good probe of magnetism in \CsE.  
\begin{figure}[b]
\centerline{\includegraphics[scale=0.47]{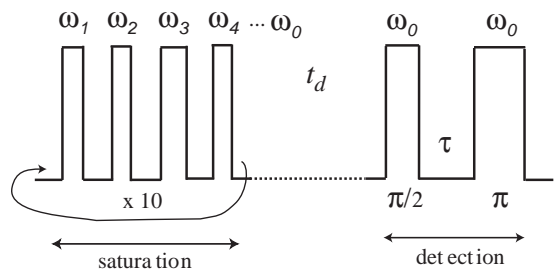}}
\caption{\label{FigSeq}   Sketch of a typical pulse sequence used to measure $T_1^{-1}$ rate.} 
\end{figure}
%

At high temperatures \To\,  is dominated by phonon contributions  \cite{Lim:2004, Abragam:1961}.
Crucial measurements for determining the value of the  spin gap are performed at lower temperatures,  
where spectra are very broad and quadrupolar satellites cannot be discerned (see \mbox{Fig. \ref{spectrum}}).
Low temperature spectra are broader than the excitation width of a typical RF pulse.
Therefore, to be able to saturate nuclear spin magnetization we had to sweep the frequency of the excitation pulses across the line. 
  More precisely, the magnetization was saturated by applying  a train of $\frac{\pi}{2}$ pulses equally spaced by a time 
$t  < T_2$ at different frequencies across the line, as illustrated in \mbox{Fig. \ref{FigSeq}}. The length of pulses was varied depending on their frequency, in order  to correctly account for the variation of $\frac{\pi}{2}$ pulse lengths   for different satellite transitions  \cite{Abragam:1961}. 
 Repeating this sequence for at least 10 times 
 ensured  a complete saturation of the nuclear magnetization. 
 Following the saturation pulse train,  the signal was detected  after a variable delay time $\tau$ using a standard 
   spin echo sequence.  The  length of the detection pulses was tuned so that only a signal from a narrow region around the  peak of the spectra, {\it i.e.},  around the central  $(+ 1/2 \leftrightarrow - 1/2)$ transition,  was detected. This procedure assured that the magnetization relaxation curves are   simple exponential functions of time. 
 %
 %
\begin{figure}[b]
\centerline{\includegraphics[scale=0.55]{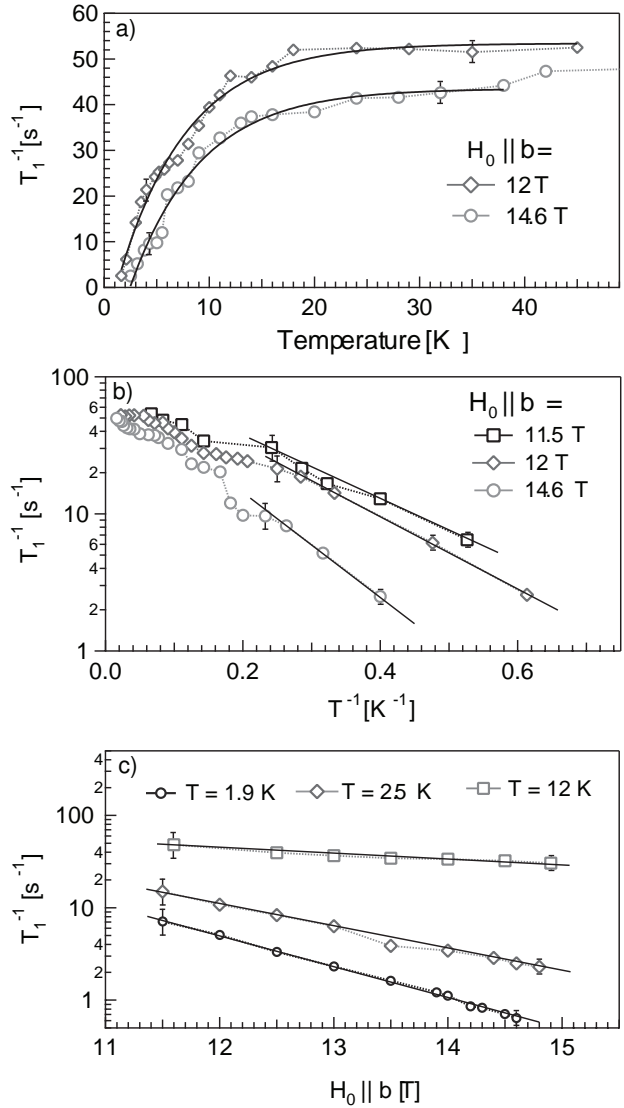}}
\caption{\label{T1}  {\bf (a)} $T_1^{-1}$ rate measurement  as a
  function of   temperature for different values of the applied
  field,  $H_0\,||\,b$. {\bf (b)} A log-inverse plot of (a) showing linear behavior at
  low temperatures. {\bf (c)} $T_1^{-1}$ as a function of the field at different 
  temperatures. The solid lines are fit to functional forms described in the text.} 
\end{figure}

 In    \mbox{Fig. \ref{T1}(a)}  the temperature  dependence   of
$T_1^{-1}$ of Cs(A)  is shown  at   $12$\,T and  $14.6$\,T  fields applied  
along the $\hat b$ axis. The rate approaches   a 
constant in the paramagnetic limit ($T\sim 50$\,K) and decreases  
exponentially at low temperature. 
The  low temperature  behavior of \To  is an indication of the opening of 
 a  gap to spin-flip  excitations
above the ferromagnetically aligned ground state.  The spin-lattice
relaxation process requires  an energy transfer of $\hbar\omega_{{\rm {\tiny NMR}}}$ of the order of 
\mbox{$1 \, {\rm mK}$} on the temperature scale. This is     negligeable compared to the spin gap 
(i.e., the minimum energy to create a magnon)
\mbox{$\Delta_{\rm H}/k_B  \sim  \, 1\, {\rm K}$}.  Thus, the relaxation
is only effective through the creation of two or more magnons in predominantly one 
 channel \cite{Beeman:1968,Sagi:1996} in the relevant field range. 
 More precisely, intrabranch transitions
involving two magnons with the same $S_z$ dominate the low temperature relaxation process. 
These have energy close to
$\Delta_{\rm H}$ and imply that the temperature dependence of the 
spin-lattice relaxation rate is given by \cite{Troyer:1994}: 
\begin{equation}
\label{T1Tdep}
T_1^{-1} \propto \exp( - \Delta_{\rm H} / k_BT),  
 \end{equation}
 where $\Delta_{\rm H}$ is defined in \mbox{Eq. (\ref{DelH})}. 
%
%

\begin{table}[b]
\caption{\label{t1table}Experimental value of the g-factor,   the saturation field, $H_{\rm sat}$, and the saturation field energy, $E_{\rm sat}$,
 at different applied field
   {\bf (a)} and temperature  {\bf (b)}.\\} 
  \begin{indented} 
\item[]
\begin{tabular}{@{}lccc}
 \hline
  {\bf (a)} & $11.5$\,T & $12$\,T & $14.6$\,T \\
  \hline
$H_{\rm sat}^{b}$\,[T]  & $8.0(1)$ & $8.2(1)$ & $8.7(4)$ \\
  \hline
  \\
  \hline
 {\bf (b)}& $1.9$\,K & $2.5$\,K & $12$\,K \\
  \hline
$g_b$  & $2.19(4)$ & $2.13(7)$ & $2.3(2)$ \\
 
$(g_b)_{\rm mean}$& \multicolumn{2}{c}{$2.16(6)$} &- \\
 
$H^{b}_{\rm sat}$\,[T] & \multicolumn{2}{c}{$8.2(1)$} & $8.7(1)$ \\
 
$E_{\rm sat}$\,[meV] & $1.04$ & $1.01$ & $1.16$ \\
\hline
 
\end{tabular}
\end{indented} 
\end{table} 
Fits to this functional form are displayed as the solid lines in  \mbox{Fig. \ref{T1}(a)}.  
Setting the $g$ factor value to  $g_b=2.103(3)$, as determined by the  ESR measurements \cite{Sharnoff:1965}   at $T=77$\,K,  we find $H_{\rm sat}$. 
 The findings are summarized in \mbox{Table \ref{t1table}(a)}.
In \mbox{Fig. \ref{T1}(b)}  the rate data is displayed on  a  log-inverse
plot. The low temperature 
  slope is given by $-g_b \mu_B (H_0-
H_{\rm sat})$.  This graph shows that up to $T \sim 5$ K,  $\ln \left(T_1^{-1}\right)$ is indeed a linear function of $T^{-1}$.

 Assuming that  the saturation field is approximately    $8.5$\,T for $H_0 || \hat{b}$, one finds that 
\mbox{$\Delta_{\rm H}/k_B = g \mu_B
(H_0 -H_{\rm sat})/k_B$} ranges from  4 K  to 8 K in the   applied  field range. This  implies that at low temperatures, $T \lesssim 5$ K, the 2-magnon intrabranch process (involving  energies of the order of $\sim 6$\,K) is 
favored compared to the other intrabranch processes, such as 3-magnon ones sensitive to   the energies of the order of 
$2 \Delta_{\rm H}/k_B \sim 12$ K.

In  \mbox{Fig. \ref{T1}(c)} $T_1^{-1}$  as a function of the applied field parallel to $\hat{b}$ axis 
at different temperatures is displayed. 
 A fit   to \mbox{Eq. (\ref{T1Tdep})}  provides a way to independently
 deduce both $H_{\rm sat}$ and $g$ values for the data. The results are  presented in  \mbox{Table \ref{t1table}(b)}. 
 From the low temperature data, $T = 1.9$   K  and 2.5  K, we find that \mbox{$g_b = 2.16(6)$} and 
 \mbox{$H_{\rm sat} = 8.2(1)$ T}. 
 The value of the saturation field   is consistent with  
the result of  the magnetization measurements \cite{Tokiwa:2005}. 
At \mbox{$T=12$\,K},
a   higher value  of \mbox{$H_{\rm sat} =8.7$\,T} is found. This is to be expected since 
at \mbox{$T=12$\,K},  \mbox{$k_BT\sim E_{\rm sat} = g\mu_B H_{\rm sat}$} and the other, mostly thermal, 
processes contribute to the rate as well.  
A relatively small discrepancy between $H_{\rm sat}$ values obtained from 12 K and 2 K data 
 indicates that the 2-magnon intrabranch processes   still 
    play a significant role in the relaxation process at 12 K.

A saturation field energy  of \mbox{$E_{\rm sat} = g\mu_B H_{\rm sat} = 1.02(2)$\,meV} is obtained for  the low temperature data. 
 The theoretical
value of the saturation field energy is given by \cite{Coldea:2002},
\begin{equation}
E_{\rm sat} = 2(J_1+J_2) + \frac{J_2^2}{2J_1}, 
\end{equation}
and is equal to \mbox{$E_{\rm sat} =1.021$\,meV} for \CsCl with 
$J_1=0.374$\,meV and $J_2=0.128$\,meV.  
Our result is  in agreement with the theoretical prediction.  For a field applied along the $\hat{a}$ axis,   the neutron experiments \cite{Coldea:2002, Radu:2005}  report  a value of
$1.065$\,meV,  consistent with our finding.

 
\section{Spin-Spin Relaxation Time}\label{T2}

The quadrupole interaction depends on the crystal structure of the
material via the electric field gradient $V_{\alpha \alpha}$. 
%
\begin{figure}[t]
\centerline{\includegraphics[scale=0.55]{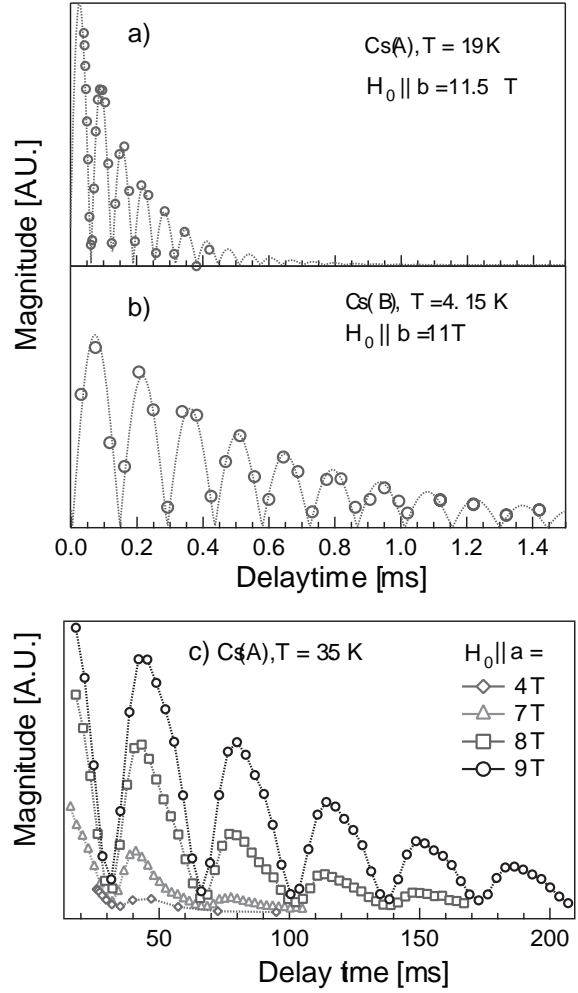}}
\caption{\label{t2} Spin echo amplitude as a function of $\tau$,
  the delay time, for site  Cs(A), {\bf (a)},  and Cs(B), {\bf (b)}. The solid lines are fit to \mbox{Eq. (\ref{bit})}. {\bf (c)}  Spin echo amplitude as a function of the applied field $H_0\,||\,a$ at
  $T=3.5$\,K for Cs(A), showing  no change of the beat frequency as $H_0$ is varied.} 
\end{figure}
%
Thus, any variations in the quadrupole splitting is a possible
indicator of structural change.  
However,  at low temperature the magnetic 
broadening of the resonance peak masks the quadrupolar splitting, as shown in \mbox{Fig. \ref{spectrum}b}. 
To assure that no structural change take place in the temperature regime of interest, 
 an alternative method to measure quadrupolar parameters is required. 
 Measurement of the spin-spin relaxation rate, $T_2^{-1}$, can be employed to infer these parameters  \cite{Abe:1966}.

  $T_2^{-1}$ is   acquired by applying
a succession of  \mbox{$\frac{\pi}{2}$-$\tau$-$\pi$-$\tau$-echo} pulses with 
 a range of values of the delay
time $\tau$. By integrating the spin-echo  as a function of the delay
time, we have obtained a succession of decreasing beats.
In \mbox{Fig.  \ref{t2}(a)},    result of a  typical $T_2$ measurement  is shown.
 The relaxation profiles can be fit to an expression given by   \cite{Abe:1966} 
\begin{equation}
\label{bit}
 M(t) = y_0 + Ae^{-(2\tau/T_2)}\cos(\omega_B \tau + \phi).
\end{equation}
where $M(t)$  is the integrated  signal magnitude, $\omega_B$  the frequency of the beats, $A$  the amplitude,  and
$\phi$ is the phase.  A net difference in the frequency
of the beats for the two sites is evident in  \mbox{Fig.  \ref{t2}(a) and \ref{t2}(b)}. In the case of Cs(B), the frequency is $\omega_B \sim
6.9(2)$\,kHz,  which is more than double the value $\omega_B \sim 16.2(5)$\,KHz found for   Cs(A).
 This suggests that the beats are caused by the quadrupole
interaction \cite{Abe:1966} and that the frequency $\omega_B$ simply equals  the quadrupole splitting frequency $\omega_Q$.
The quadrupolar 
splitting frequencies inferred from the beats in the \Tt\, relaxation profile equal,  within  error,  
the  $\omega_Q$ measured directly from the satellite splitting at high temperatures (see Section III).
More importantly, we find the same beat values in all fields and at all temperatures down to $2$\,K. 
\mbox{Fig. \ref{t2}(c)} shows results for Cs(A) at $T=3.5$\,K for different fields applied along the $\hat{a}$ axis.
We clearly see that the frequency of the beats is unchanged as the applied  field is  increased. 
Therefore, we  conclude that no significant changes in the crystal structure of the material take 
place in the temperature and field regime investigated in this work. 
 Finally, the data displayed in \mbox{Fig. \ref{t2}(c)} indicate  that   the relaxation time $T_2$ increases  as 
 the field is increased. This is expected since a stronger field   freezes the magnetic fluctuations,  and therefore slows down the processes that cause spin decoherence.

\section{Conclusion}

An extensive $^{133}$Cs NMR study  
performed on the 2D frustrated Heisenberg antiferromagnet
Cs$_2$CuCl$_4$ at temperatures down to $2$\,K and in a magnetic field up to $15$\,T, is reported. 
Spectra are
characterised by two main resonance peaks from two inequivalent cesium sites, referred to as  Cs(A) and Cs(B). 
The net shifts
have been deduced for a wide range of fields and temperatures. From
these results, we have concluded that the local field is due to a
strong transfer hyperfine interaction between the Cu spins and the Cs nuclei.  
The hyperfine coupling constants have been deduced. They are strongly anisotropic for   
   the Cs(B) site.
 From the spin-lattice relaxation rate measurements   in  fields higher than the
saturation field,  we inferred 
the values of the $g$-factor and the
saturation field $H_{\rm sat}$ along the $\hat b$ axis.  These are in agreement with the 
  neutron scattering and magnetization measurement results. Spin-spin relaxation rate measurements 
are used to deduce quadrupolar splitting frequencies and thus assure that no structural 
phase transitions take place at temperatures as low as $1.9$\,K and in magnetic fields up to $15$\,T.

Our work demonstrates that \Cs NMR is a good probe of the Cu spin degrees of freedom. 
The measurements are necessary precursory work 
 to an investigation of the intriguing low temperature magnetic phases. In these phases  previous neutron scattering 
 \cite{Coldea:2003} and magnetization measurements \cite{Tokiwa:2005} have shown evidence 
 for correlated quantum fluctuations enhanced by the low spin and frustrated triangular geometry.

 \ack{
We are very grateful to J. B. Marston and S. Ma for helpful discussions.   
The work was supported in part by the 
the National Science Foundation    (DMR-0547938) and  the funds from the Brown University and Salomon Research Fund. 
The work at the National High
Magnetic Field Laboratory was supported by the National Science Foundation 
under Cooperative Agreement No. DMR95-27035 and the State of Florida.}

\vspace{0.5cm}


\begin{thebibliography}{10}
\bibitem{NotPolField}
   Precise values of $H^{\rm{int}}$ was determined from the shift of the NMR line measured at low temperatures ($T < 100$ mK) and high fields ($H_0 > 12$ T).
\end{thebibliography}
\end{document}